# Transferability of interatomic potentials for germanene (2D germanium)

Marcin Maździarz[a]
*Institute of Fundamental Technological Research Polish Academy of Sciences, Pawińskiego 5B, 02-106 Warsaw, Poland*

The capacities of various interatomic potentials available for elemental germanium, with the scope to choose the potential suitable for the modeling of germanene (2D germanium) allotropes were investigated. Structural and mechanical properties of the flat (F), low-buckled (LB), trigonal dumbbell (TD) and large honeycomb dumbbell (LHD) single-layer germanium (germanene) phases, were obtained using the density functional theory (DFT) and molecular statics (MS) computations with Tersoff, MEAM, Stillinger-Weber, EDIP, ReaxFF and machine-learning-based (ML-IAP) interatomic potentials. A systematic quantitative comparative study and discussion of the findings are given.

## I. INTRODUCTION

The fabrication of graphene, a flat carbon structure, in 2004[1] and then, in 2010, the awarding of this achievement with the Nobel Prize in Physics sparked a huge global interest in 2D materials. The Computational 2D Materials Database currently lists more than 4000 hypothetical two-dimensional (2D) materials[2]. Graphene is extremely strong, thin, flexible and lightweight. It conducts electricity and heat exceptionally well. Unfortunately, it is not a semiconductor, and cannot be switched off. Natural candidates for flat semiconductors were other chemical elements from the carbon group of the 14th (formerly IVA or IV main) group of the periodic table, i.e. silicon (Si), germanium (Ge), tin (Sn), lead (Pb) and flerovium (Fl).

Germanene is an allotropic form of germanium (Ge, Latin: *germanium*), a nanomaterial with a flat structure analogous to graphene (2D carbon), silicene (2D silicon), stanene (2D tin) and plumbene (2D lead), all already synthesized[3]. Silicon and germanium have very similar physical properties, this also applies to their 2D allotropes.

Although the first transistors were made from germanium (Ge) and the potential barrier of silicon (Si) is greater than that of germanium, nowadays silicon dominates the semiconductor industry. Simply put, Si has better temperature stability, smaller variation of leakage current, and lower cost. Since the importance of germanium is much lower compared to silicon, this is manifested both in the interest of researchers and in such an issue as the number of available interatomic potentials, see NIST Interatomic Potentials Repository[4] and OpenKIM Repository[5]. However, some augur a renaissance of germanium[6].

Analyzing the available literature, we can find up to ten hypothetical thin allotropes of germanium[7,8]. In the present work, we deal with 2D single-layer (SL) germanium and some of these proposed exotic thin allotropes of Ge, such as kagome, bridge or MoS$_2$-like lattice can hardly be considered as SL structures, they are simply two and three-layered. The available interatomic potentials for germanium were not designed with its 2D allotropes in mind. In the papers where they were proposed, they were tested and compared with each other for data for 3D germanium. The exception is a paper where the Tersoff potential is re-fitted to the structural and mechanical properties of the low-buckled germanene phase[9]. In order to test the versatility, transferability of these potentials, four 2D SL germanium allotropes were finally used, i.e. flat (F), low-buckled (LB), trigonal dumbbell (TD) and large honeycomb dumbbell (LHD).

---

[a] Electronic mail: mmazdz@ippt.pan.pl

The problem of transferability/universality of interatomic potentials to systems outside the training set has also attracted the attention of other researchers, e.g. the ML-IAP potentials for zinc oxide polymorphs have been studied in[10], the embedded-atom-method (EAM) based potentials for some transition metals in[11], various potentials for disordered carbon structures in[12], and the potentials available for elemental carbon to model penta-graphene, a two-dimensional carbon allotrope, in[13].

To the best of the author's knowledge, there is no publication that evaluates the ability of interatomic potentials for germanium to describe its 2D allotropes, nor a publication that comprehensively analyses the dynamical and mechanical stability of germanene allotropes using *ab initio* calculations. The aim of this work is first to determine the structural and mechanical characteristics of two-dimensional germanium polymorphs using the *ab initio* methods, and then to test the capability of available interatomic potentials to reproduce these properties.

## II. METHODS

Since the importance in industry, economy of germanium is much smaller than silicon then also the availability of research on it is reduced, this also applies to its 2D single-layer (SL) allotropes. If even data are available it is highly incomplete and obtained with different methodologies. The first part of the study was to determine in one consistent way, using *ab initio* calculations described in the Section II A, structural and mechanical data for the four 2D single-layer germanium polymorphs analyzed. These data include lattice parameters, average cohesive energy, average bond length, average height, 2D elastic constants as well as phonon spectrum. The data so determined were further treated as reference data and labeled as Value$^{DFT}$. The same data were then calculated using classical molecular statics (MS), see the Section II A, and the twelve interatomic potentials analyzed from Subsection II B. These results were labeled as Value$^{potential}$. Having reference data and those from MS calculations, we define the mean absolute percentage error (MAPE):

$$MAPE = \frac{100\%}{n} \sum_{t=1}^{n} \left| \frac{Value^{DFT} - Value^{Potential}}{Value^{DFT}} \right| \quad (1)$$

that enables us to quantify the interatomic potentials being tested.

Additionally, a series of molecular dynamics (MD) simulations (200 atoms and 10000 time steps, NVE-microcanonical ensemble) and the built-in LAMMPS function *timesteps/s* were utilized to evaluate the computational cost of the analyzed interatomic potentials. The results were then normalized against the longest simulation time. (For performance testing, the 1-processor runs were carried out on a single core of an Intel Xeon Platinum 8268 cluster system, running Linux (GNU Linux, 64-bit). LAMMPS (version 2 August 2023) was compiled using the GCC compiler, version 12.2.)

### A. Ab Initio and Molecular Calculations

The *ab initio* (structures' relaxation, calculation of cohesive energies, 2D elastic constants and phonons), and molecular calculation (structures' relaxation, calculation of cohesive energies and 2D elastic constants, CPU cost of various potentials) methodology here is directly borrowed from[14]. The programs used here are: for the density functional theory (DFT)[15,16] and the density functional perturbation theory (DFPT)[17] ABINIT[18], for the molecular statics (MS) and the molecular dynamics (MD) the Large-scale Atomic/Molecular Massively

Parallel Simulator (LAMMPS)[19], for visualization and analysis the Open Visualization Tool (OVITO)[20].

B.  Interatomic Potentials

The parametrizations of the potentials enumerated below were obtained from the NIST Interatomic Potentials Repository and/or from LAMMPS code sources and/or thanks to the goodwill of the authors of the publications.

1. Tersoff1989[21]: the Tersoff potential fitted to polytype energies of 3D germanium

2. Tersoff2017[9]: a re-parametrization of the Tersoff potential fitted to structural and mechanical properties of low-buckled germanene

3. MEAM2008[22]: the modified embedded atom method (MEAM) potential fitted to elastic, structural, point defect, surface and thermal properties of 3D germanium

4. SW1986[23]: the Stillinger–Weber (SW) potential fitted to cohesive energy and elastic constants of 3D germanium

5. SW2009[24]: the Stillinger–Weber (SW) potential fitted to cohesive energy and lattice constant of diamond-structure germanium

6. EDIP[25]: the environment-dependent interatomic potential (EDIP) for 3D germanium fitted to *ab initio* results published in the literature

7. ReaxFF[26]: the reactive force-field (ReaxFF) fitted to a training set of *ab initio* data that pertain to Si/Ge/H bonding environments

8. SNAP2020[27]: the machine-learning-based (ML-IAP) linear variant of the spectral neighbor analysis potential (SNAP) fitted to total energies and interatomic forces in ground-state 3D Ge, strained structures and slab structures obtained from *ab initio* computations

9. qSNAP2020[27]: the machine-learning-based (ML-IAP) quadratic variant of the spectral neighbor analysis potential (qSNAP) fitted to total energies and interatomic forces in ground-state 3D Ge, strained structures and slab structures obtained from *ab initio* computations

10. ACE[28]: the machine-learning-based (ML-IAP) variant of the atomic cluster expansion potential (ACE) fitted to previously published *ab initio* data sets for 3D germanium

11. SNAP2023[28]: newer parametrization of the machine-learning-based (ML-IAP) linear variant of spectral neighbor analysis potential (SNAP) fitted to previously published *ab initio* data sets for 3D germanium

12. POD[28]: the proper orthogonal descriptors based potential (POD) fitted to previously published *ab initio* data sets for 3D germanium

III.  RESULTS

The first stage of *ab initio* calculations was an attempt to relax/optimize 6 hypothetical 2D structures already reported for silicon[14]. The Si atoms were simply replaced by Ge atoms. It was found that during the relaxation/optimization of the structures, the high-buckled (HB) converged to the low-buckled LB[29], and the initial honeycomb dumbbell (HD) did not converge at all in 100 steps. The resulting unit cells for the four germanene allotropes, i.e, the flat (F):(*hP2*, P6/mmm, No.191), low-buckled (LB):(*hP2*, P$\bar{3}$m1, No.164), trigonal dumbbell (TD):(*hP7*, P$\bar{6}$2m, No.189) and large honeycomb dumbbell (LHD):(*hP10*, P6/mmm, No.191) are illustrated in Figs. 1a)-d) and the extra crystallographic data for them are retained in the Crystallographic Information Files (CIFs) in the Supplementary material V. All the germanene allotropes analyzed here similarly to silicene have hexagonal symmetry, this should imply isotropy of physical properties in the plane[14].

The phonon spectra along the Γ-M-K-Γ path for the four germanene allotropes is illustrated in Figs. 2a)-d). The study of the calculated curves indicates that the LB and LHD phases are dynamically stable, i.e., there are no phonon modes with negative frequencies anywhere. The flat (F) phase and trigonal dumbbell (TD) can theoretically be dynamically unstable, i.e., some phonon modes have a negative frequency. A flat free-standing germanene has not been noted in observations, however, it has been observed on a suitable substrate[30]. It was therefore decided that these two phases would also be included in the molecular computations.

Structural and mechanical properties of the four germanene allotropes determined from *ab initio* calculations, namely lattice parameters, average cohesive energy, average bond length, average height, 2D elastic constants, 2D Young's modulus, Poisson's ratio and 2D Kelvin moduli are gathered in Table I. Data available from other calculations for LB and LHD germanene allotropes are in reasonable agreement with the present results. It is immediately notable that all the computed 2D Kelvin moduli for all the germanene phases are positive, resulting in mechanical stability[31].

The computed with the use of molecular statics and the twelve various interatomic potentials for germanium (Tersoff (×2), MEAM, Stillinger-Weber (×2), EDIP, ReaxFF and machine-learning-based (ML-IAP (×5)), enumerated in Section II B, twelve structural and mechanical properties, i.e., lattice parameters a, b, average cohesive energy $E_c$, average bond length $d$, average height $h$, 2D elastic constants $C_{ij}$, 2D Kelvin moduli $K_i$ of the flat germanene (F) phase are collected in Table II, of the low-buckled germanene (LB) in Table III, of the trigonal dumbbell germanene (TD) in Table IV and of the large honeycomb dumbbell germanene (LHD) in Table V, respectively. The above mentioned results, for each of the four germanene allotropes, were then compared with those from the DFT calculations using the mean absolute percentage error (MAPE) defined in Eq. 1. Looking at the results for each phase, we see that for the F phase the most accurate are the ReaxFF and Tersoff1989 potentials, they have the lowest MAPE, see Table II, for the LB phase the ReaxFF, see Table III, for the TD phase the Tersoff1989, see Table IV and for the LHD the ReaxFF and Tersoff2017 potentials, see Table V, respectively. When we look summarily we see that only the EDIP potential is not able to correctly reproduce the structural properties of the four polymorphs of germanene, see Tabs. II. Although the ACE potential correctly reconstructs the geometry of all polymorphs, it has problems in reproducing the required isotropy of the stiffness tensor, see Tables II-V. The ReaxFF potential spuriously shows a lack of mechanical stability for the TD phase, see Table IV. It seems that ACE potential is incorrectly implemented in the LAMMPS code (verified on both Windows and Linux), or simply doesn't work properly for 2D structures. For ReaxFF potential, the deviation from isotropy occurs only for the LHD phase, is small, and is most likely due to problems with the charge equilibration (QEq) method. The SNAP2020 potential gives nonsensical zero elastic constants for the LHD phase, see Table V. This seems to be caused by the oversized lattice constants predicted by this potential.

Among these twelve potentials, two: ReaxFF and SW1986 provide the best quantitative performance measured by the total mean absolute percentage error (MAPE), see Table V. Looking at the cost of computations in terms of relative performance measured as normalized timesteps/second in molecular dynamics (MD) simulations the EDIP potential is the fastest, about 100 times faster than ReaxFF, and up to 500 times faster than the ML-IAP(SNAP2020) potential, see Table V. The results obtained here confirm the previous observations[14,35] that the machine-learning-based (ML-IAP) interatomic potentials, in accordance with the methodology applied, are not superior to classical potentials in terms of their performance/accuracy (MAPE) and are meanwhile up to two orders of magnitude more computationally expensive, see Table V.

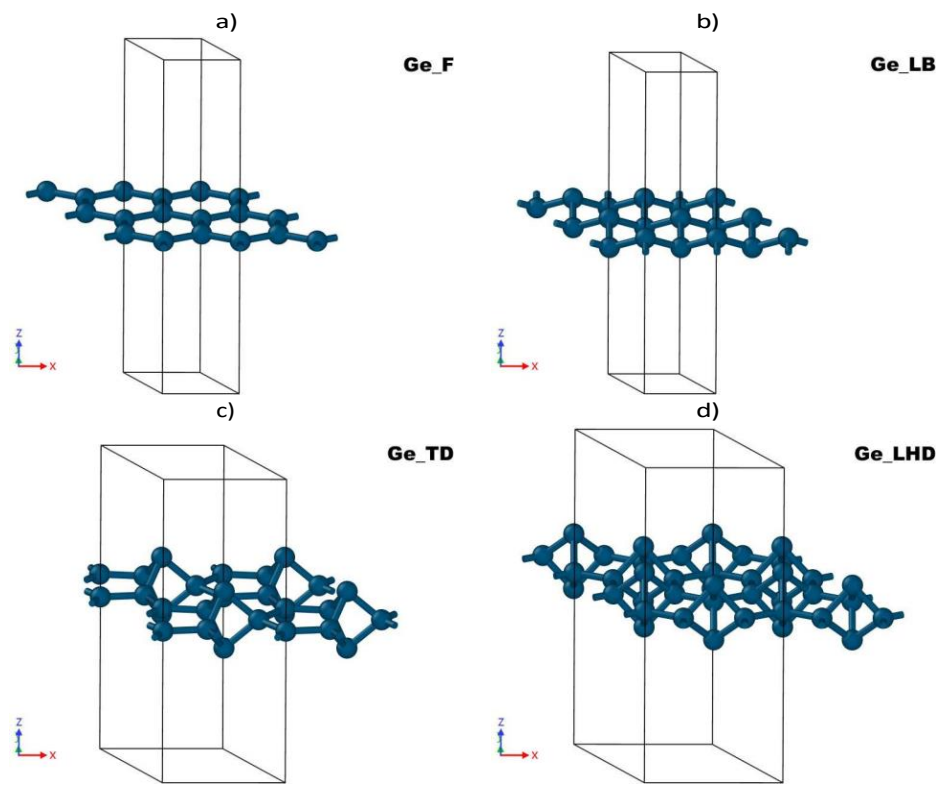

FIG. 1. Polymorphs of germanene: a) flat (F), b) low-buckled (LB), c) trigonal dumbbell (TD), d) large honeycomb dumbbell (LHD).

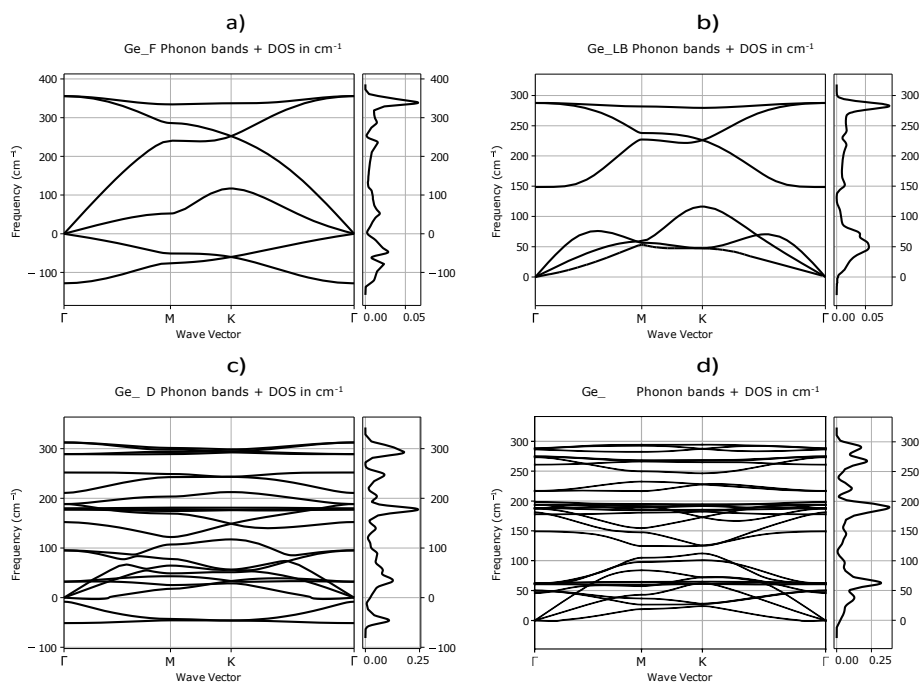

FIG. 2. Phonon dispersion and density of states (DOS) of the a) flat (F), b) low-buckled (LB), c) trigonal dumbbell (TD), d) large honeycomb dumbbell (LHD) single-layer germanium (germanene) phases. High symmetry points: Γ[0,0,0], M[1/2,0,0], K[1/3,1/3,0].

TABLE I. Structural and mechanical properties of flat (F), low-buckled (LB), trigonal dumbbell (TD) and large honeycomb dumbbell (LHD) germanene phases from density functional theory (DFT) calculations: lattice parameters a,b (Å), average cohesive energy $E_c$ (eV/atom), average bond length $d$ (Å), average height $h$ (Å), 2D elastic constants $C_{ij}$ (N/m), 2D Young's modulus E (N/m), Poisson's ratio $v$ and 2D Kelvin moduli $K_i$ (N/m).

| Polymorph | F | | LB | | TD | | LHD | |
|---|---|---|---|---|---|---|---|---|
| Source | This work | Refs. | This work | Refs. | This work | Refs. | This work | Refs. |
| a | 4.066 | | 3.955 | 4.06[a], 4.03[b], 3.97[c] | 6.698 | 6.772[a] | 7.667 | 7.88[a] |
| b | 4.066 | | 3.955 | 4.06[a], 4.03[b], 3.97[c] | 6.698 | 6.772[a] | 7.667 | 7.88[a] |
| $-E_c$ | 3.566 | | 3.887 | 3.19 , 4.15 | 3.793 | 3.30 | 3.863 | 3.35 |
| $d$ | 2.349[†] | | 2.370 | 2.47[b], 2.38[c] | 2.449 | | 2.485 | |
| $h$ | 0.0 | | 0.636 | 0.69 , 0.68 , 0.64[c] | 2.977 | 2.90[a] | 2.875 | 2.96[d] |
| $C_{11}$ | 70.34 | | 49.86 | | 53.56 | | 57.72 | 48.9[d] |
| $C_{22}$ | 70.34 | | 49.86 | | 53.56 | | 57.72 | 48.9[d] |
| $C_{12}$ | 29.64 | | 15.56 | | 17.20 | | 20.68 | |
| $C_{33}$ | 20.35 | | 17.15 | | 18.18 | | 18.52 | |
| E | 57.85 | | 45.00 | 42.05[b] | 48.04 | | 50.31 | |
| $v$ | 0.42 | | 0.31 | 0.33[b] | 0.32 | | 0.36 | |
| $K_I$ | 99.98 | | 65.42 | | 70.76 | | 78.40 | |
| $K_{II}$ | 40.70 | | 34.30 | | 36.36 | | 37.04 | |
| $K_{III}$ | 40.70 | | 34.30 | | 36.36 | | 37.04 | |

[a] Ref. [7], [b] Ref. [32], [c] Ref. [33], [d] Ref. [34].
[†] An average bond lengths calculated using radial pair distribution function with a *cut-off* radius = 3.5 Å and a number of histogram bins = 1000[20].

TABLE II. Structural and mechanical properties of flat (F) germanene from molecular calculations: lattice parameters a, b (Å), average cohesive energy $E_c$ (eV/atom), average bond length $d$ (Å), average height $h$ (Å), 2D elastic constants $C_{ij}$ (N/m), 2D Kelvin moduli $K_i$ (N/m), mean absolute percentage error (MAPE) (%).

| Method | DFT | Tersoff 1989 | Tersoff 2017 | MEAM 2008 | SW 1986 | SW 2009 | EDIP | ReaxFF | qSNAP 2020 | SNAP 2020 | SNAP 2023 | ACE | POD |
|---|---|---|---|---|---|---|---|---|---|---|---|---|---|
| a | 4.066 | 4.222 | 4.190 | 3.955 | 4.287 | 4.268 | 4.627 | 4.175 | 4.678 | 4.351 | 4.419 | 4.345 | 4.195 |
| b | 4.066 | 4.222 | 4.190 | 3.955 | 4.287 | 4.268 | 4.627 | 4.175 | 4.678 | 4.351 | 4.419 | 4.345 | 4.195 |
| $-E_c$ | 3.566 | 2.974 | 3.061 | 2.606 | 2.755 | 2.702 | 4.665 | 5.780 | 4.452 | 4.382 | 4.042 | 3.887 | 4.274 |
| d | 2.349 | 2.440 | 2.417 | 2.286 | 2.475 | 2.475 | 2.671 | 2.412 | 2.699 | 2.510 | 2.552 | 2.510 | 2.426 |
| h | 0.0 | 0.0 | 0.0 | 0.0 | 0.0 | 0.0 | 0.0 | 0.0 | 0.0 | 0.0 | 0.0 | 0.0 | 0.0 |
| $C_{11}$ | 70.34 | 59.76 | 50.46 | 103.22 | 50.57 | 42.05 | 45.02 | 65.05 | 44.87 | 32.44 | 22.60 | 30.04 | 75.75 |
| $C_{22}$ | 70.34 | 59.76 | 50.46 | 103.22 | 50.57 | 42.05 | 45.02 | 65.05 | 44.87 | 32.44 | 22.60 | 30.04 | 75.75 |
| $C_{12}$ | 29.64 | 16.68 | 28.47 | 44.15 | 25.22 | 32.47 | 11.14 | 29.65 | 28.62 | 18.04 | 17.52 | 13.03 | 68.35 |
| $C_{33}$ | 20.35 | 21.54 | 10.99 | 29.53 | 12.67 | 4.79 | 16.94 | 17.70 | 8.12 | 7.20 | 2.54 | 5.13 | 3.70 |
| $K_I$ | 99.98 | 76.46 | 78.93 | 147.37 | 75.78 | 74.52 | 56.16 | 94.71 | 73.49 | 50.48 | 40.12 | 43.07 | 144.10 |
| $K_{II}$ | 40.70 | 43.07 | 21.99 | 59.07 | 25.35 | 9.59 | 33.88 | 35.40 | 16.24 | 14.40 | 5.07 | 17.01 | 7.40 |
| $K_{III}$ | 40.70 | 43.07 | 21.99 | 59.07 | 25.35 | 9.59 | 33.88 | 35.40 | 16.24 | 14.40 | 5.07 | 17.01 | 7.40 |
| MAPE$_F$ |  | 13.00 | 22.06 | 32.76 | 22.49 | 34.93 | 27.32 | 11.78 | 32.04 | 39.45 | 48.95 | 42.26 | 42.28 |

TABLE III. Structural and mechanical properties of low-buckled (LB) germanene from molecular calculations: lattice parameters a, b (Å), average cohesive energy $E_c$ (eV/atom), average bond length $d$ (Å), average height $h$ (Å), 2D elastic constants $C_{ij}$ (N/m), 2D Kelvin moduli $K_i$ (N/m), mean absolute percentage error (MAPE) (%).

| Method | DFT | Tersoff 1989 | Tersoff 2017 | MEAM 2008 | SW 1986 | SW 2009 | EDIP | ReaxFF | qSNAP 2020 | SNAP 2020 | SNAP 2023 | ACE | POD |
|---|---|---|---|---|---|---|---|---|---|---|---|---|---|
| a | 3.955 | 4.125 | 3.950 | 3.862 | 3.998 | 4.012 | 4.627 | 3.978 | 4.560 | 3.653 | 4.212 | 4.046 | 4.040 |
| b | 3.955 | 4.125 | 3.950 | 3.862 | 3.998 | 4.012 | 4.627 | 3.978 | 4.560 | 3.653 | 4.212 | 4.046 | 4.040 |
| $-E_c$ | 3.687 | 3.016 | 3.181 | 2.716 | 2.895 | 2.740 | 4.665 | 6.085 | 4.475 | 4.464 | 4.054 | 4.010 | 4.288 |
| d | 2.370 | 2.433 | 2.404 | 2.328 | 2.447 | 2.454 | 2.671 | 2.447 | 2.692 | 2.517 | 2.538 | 2.524 | 2.440 |
| h | 0.636 | 0.491 | 0.760 | 0.665 | 0.816 | 0.819 | 0.000[‡] | 0.845 | 0.557 | 1.369 | 0.714 | 0.960 | 0.708 |
| $C_{11}$ | 49.86 | 93.39 | 94.59 | 79.48 | 38.92 | 14.14 | 45.02 | 48.16 | 19.11 | 34.55 | 10.63 | 26.93[†] | 31.42 |
| $C_{22}$ | 49.86 | 93.39 | 94.59 | 79.48 | 38.92 | 14.14 | 45.02 | 48.16 | 19.11 | 34.55 | 10.63 | 25.14[†] | 31.42 |
| $C_{12}$ | 15.56 | -20.29 | -16.14 | 16.15 | 4.50 | 3.61 | 11.14 | 13.76 | 7.78 | 1.55 | 1.04 | 11.49[†] | 5.93 |
| $C_{33}$ | 17.15 | 56.84 | 55.36 | 31.67 | 17.21 | 5.27 | 16.94 | 17.20 | 5.67 | 16.50 | 4.80 | 4.77[†] | 12.75 |
| $K_I$ | 65.42 | 113.67 | 110.72 | 95.63 | 43.42 | 17.75 | 56.16 | 61.92 | 26.89 | 36.10 | 11.67 | 37.49 | 37.35 |
| $K_{II}$ | 34.30 | 73.10 | 78.45 | 63.33 | 34.42 | 10.53 | 33.88 | 34.40 | 11.34 | 32.99 | 9.59 | 14.58 | 25.49 |
| $K_{III}$ | 34.30 | 113.67 | 110.72 | 63.33 | 34.42 | 10.53 | 33.88 | 34.40 | 11.34 | 32.99 | 9.59 | 9.53 | 25.49 |
| MAPE$_{LB}$ |  | 92.25 | 88.46 | 38.34 | 17.07 | 46.80 | 19.90 | 10.57 | 42.59 | 30.44 | 49.27 | 36.42 | 24.22 |

[†] Potential does not reproduce the isotropy of the elasticity tensor ($C_{11}$ /= $C_{22}$ and 2·$C_{33}$ /= $C_{11} - C_{12}$),
[‡] Input low-buckled (LB) structure converges to flat (F) one.

TABLE IV. Structural and mechanical properties of trigonal dumbbell (TD) germanene from molecular calculations: lattice parameters a, b (Å), average cohesive energy $E_c$ (eV/atom), average bond length $d$ (Å), average height $h$ (Å), 2D elastic constants $C_{ij}$ (N/m), 2D Kelvin moduli $K_i$ (N/m), mean absolute percentage error (MAPE) (%).

| Method | DFT | Tersoff 1989 | Tersoff 2017 | MEAM 2008 | SW 1986 | SW 2009 | EDIP | ReaxFF | qSNAP 2020 | SNAP 2020 | SNAP 2023 | ACE | POD |
|---|---|---|---|---|---|---|---|---|---|---|---|---|---|
| a | 6.698 | 6.904 | 6.738 | 6.670 | 6.933 | 6.567 | 7.859 | 6.726 | 7.301 | 7.291 | 7.346 | 7.019 | 7.043 |
| b | 6.698 | 6.904 | 6.738 | 6.670 | 6.933 | 6.567 | 7.859 | 6.726 | 7.301 | 7.291 | 7.346 | 7.019 | 7.043 |
| $-E_c$ | 3.793 | 2.961 | 3.277 | 2.731 | 2.817 | 3.108 | 5.094 | 3.060 | 4.358 | 4.512 | 4.061 | 4.050 | 4.505 |
| d | 2.449 | 2.532 | 2.577 | 2.486 | 2.596 | 2.584 | 2.793 | 2.532 | 2.532 | 2.600 | 7.142 | 2.552 |  |
| h | 2.977 | 3.097 | 3.098 | 3.284 | 3.457 | 3.862 | 2.955 | 3.227 | 3.931 | 2.478 | 2.710 | 3.250 | 2.954 |
| $C_{11}$ | 53.56 | 57.37 | 51.46 | 88.32 | 70.50 | 44.75 | 30.10 | 36.17 | 22.27 | 37.51 | 18.43 | 29.95[†] | 31.67 |
| $C_{22}$ | 53.56 | 57.37 | 51.46 | 88.32 | 70.50 | 44.75 | 30.10 | 36.17 | 22.27 | 37.51 | 18.43 | 96.70[†] | 31.67 |
| $C_{12}$ | 17.20 | 16.77 | 27.47 | 31.09 | 29.02 | 30.20 | 16.66 | 35.90 | 9.99 | 21.85 | 11.32 | -1.37[†] | 10.85 |
| $C_{33}$ | 18.18 | 20.30 | 12.00 | 28.62 | 20.74 | 7.27 | 6.72 | -0.36 | 6.14 | 7.83 | 3.55 | 9.01[†] | 10.41 |
| $K_I$ | 70.76 | 74.14 | 78.93 | 119.42 | 99.52 | 74.95 | 46.76 | 72.07 | 32.26 | 59.36 | 29.75 | 96.73 | 42.51 |
| $K_{II}$ | 36.36 | 40.60 | 23.99 | 57.23 | 41.49 | 14.55 | 13.43 | 0.26 | 12.28 | 15.65 | 7.11 | 29.92 | 20.82 |
| $K_{III}$ | 36.36 | 40.60 | 23.99 | 57.23 | 41.49 | 14.55 | 13.43 | -0.72[*] | 12.28 | 15.65 | 7.11 | 18.01 | 20.82 |
| MAPE$_{TD}$ |  | 7.66 | 17.10 | 41.02 | 22.48 | 29.30 | 33.13 | 42.77 | 40.91 | 27.57 | 42.18 | 50.43 | 26.74 |

[†] Potential does not reproduce the isotropy of the elasticity tensor ($C_{11}$ /= $C_{22}$ and 2·$C_{33}$ /= $C_{11} - C_{12}$),
[*] Negative Kelvin moduli $K_i$ indicating a lack of mechanical stability.

TABLE V. Structural and mechanical properties of large honeycomb dumbbell (LHD) germanene from molecular calculations: lattice parameters a, b (Å), average cohesive energy $E_c$ (eV/atom), average bond length $d$ (Å), average height $h$ (Å), 2D elastic constants $C_{ij}$ (N/m), 2D Kelvin moduli $K_i$ (N/m), mean absolute percentage error (MAPE) (%), relative performance measured as normalized timesteps/second in molecular dynamics (MD) simulation.

| Method | DFT | Tersoff 1989 | Tersoff 2017 | MEAM 2008 | SW 1986 | SW 2009 | EDIP | ReaxFF | qSNAP 2020 | SNAP 2020 | SNAP 2023 | ACE | POD |
|---|---|---|---|---|---|---|---|---|---|---|---|---|---|
| a | 7.667 | 7.763 | 7.527 | 7.517 | 7.791 | 7.243 | 9.018 | 7.660 | 8.702 | 11.746 | 8.319 | 7.879 | 8.184 |
| b | 7.667 | 7.763 | 7.527 | 7.517 | 7.791 | 7.243 | 9.018 | 7.660 | 8.702 | 11.746 | 8.319 | 7.879 | 8.184 |
| $-E_c$ | 3.863 | 2.960 | 3.374 | 2.817 | 2.844 | 3.277 | 5.272 | 3.152 | 4.544 | 5.682 | 4.070 | 4.117 | 4.583 |
| $d$ | 2.485 | 2.560 | 2.617 | 2.700 | 2.642 | 2.617 | 2.838 | 2.664 | 2.682 | 2.220 | 2.615 | 2.746 | 2.600 |
| $h$ | 2.875 | 3.581 | 3.100 | 3.831 | 3.449 | 3.852 | 2.966 | 3.229 | 2.741 | 3.636 | 2.708 | 3.273 | 2.787 |
| $C_{11}$ | 57.72 | 55.39 | 54.55 | 78.50 | 76.29 | 46.06 | 24.56 | 56.43[†] | 24.17 | 0.00[*] | 19.34 | 53.99[†] | 48.20 |
| $C_{22}$ | 57.72 | 55.39 | 54.55 | 78.50 | 76.29 | 46.06 | 24.56 | 51.68[†] | 24.17 | 0.00[*] | 19.34 | 22.30[†] | 48.20 |
| $C_{12}$ | 20.68 | 18.62 | 25.44 | 11.54 | 35.40 | 32.05 | 17.71 | 14.90[†] | 9.84 | 0.00[*] | 10.85 | 33.11[†] | 28.04 |
| $C_{33}$ | 18.52 | 18.39 | 14.56 | 33.48 | 20.44 | 7.01 | 3.44 | 16.29[†] | 7.17 | 0.00[*] | 4.24 | 13.59[†] | 10.08 |
| $K_I$ | 78.40 | 74.01 | 79.99 | 90.04 | 111.69 | 78.10 | 42.27 | 68.77 | 34.01 | 0.00 | 30.19 | 67.22 | 76.25 |
| $K_{II}$ | 37.04 | 36.77 | 29.11 | 66.95 | 40.89 | 14.01 | 6.86 | 39.35 | 14.34 | 0.00 | 8.49 | 9.07 | 20.16 |
| $K_{III}$ | 37.04 | 36.77 | 29.11 | 66.95 | 40.89 | 14.01 | 6.86 | 32.58 | 14.34 | 0.00 | 8.49 | 27.17 | 20.16 |
| M$\Sigma$APE$_{LHD}$ | | 26.98 | 10.80 | 27.02 | 22.09 | 28.98 | 42.40 | 10.11 | 38.87 | 74.22 | 42.23 | 25.62 | 20.66 |
| MAPE | | 139.88 | 138.42 | 139.14 | 84.13 | 140.02 | 122.76 | 75.22 | 154.41 | 171.68 | 182.63 | 154.73 | 113.89 |
| timesteps/s | | 120.05 | 120.07 | 194.94 | 167.50 | 150.19 | 524.37 | 6.75 | 2.00 | 1.00 | 3.35 | 6.94 | 1.79 |

[†] Potential does not reproduce the isotropy of the elasticity tensor ($C_{11} \neq C_{22}$ and $2 \cdot C_{33} \neq C_{11} - C_{12}$),

[*] Nonsensical zero 2D elastic constants $C_{ij}$.

IV. CONCLUSION

In this paper, a systematic quantitative comparative study of different interatomic potentials of germanium was carried out to reproduce the properties of four allotropes of germanene (2D germanium). To compare the twelve potentials enumerated in Section II B, the structural and mechanical properties of flat (F), low-buckled (LB), trigonal dumbbell (TD) and large honeycomb dumbbell (LHD) germanene (Figs. 1a)-d)) obtained through density functional theory (DFT) and molecular statics (MS) calculations were utilized. The computational cost and performance of the analyzed potentials were also benchmarked.

- Only the EDIP potential is unable to correctly reproduce the structural properties of the four polymorphs of germanene, see Table III;

- The ReaxFF, ACE, and SNAP2023 potentials do not do well in describing the mechanical properties of the four allotropes of germanene;

- Two potentials: ReaxFF and SW1986 give the best quantitative performance measured by the total mean absolute percentage error (MAPE), see Table V;

- Machine-learning-based (ML-IAP) interatomic potentials, according to the methodology used, are not superior in performance (MAPE) to classical potentials and are instead even two orders of magnitude more computationally expensive, see Table V;

- Taking into account the performance/accuracy and cost of computation, the classical potentials of Tersoff, MEAM and SW type seem to be the best choice here. Although the data for various germanene allotropes were not used in the optimization of these potentials, they were capable of reproducing their properties well. This is a consequence of the fact that they are based on physical grounds, have a natural extrapolation capability, and do not just interpolate data.

I hope that the observations made here will help other researchers in selecting the right potentials for their purposes and will guide the parameterization of new potentials for germanene.

## V. SUPPLEMENTARY MATERIAL

Crystallographic Information Files (CIFs) for polymorphs of germanene (created by qAgate: Open-source software to post-process ABINIT) are available online at Supplementary material.

## ACKNOWLEDGMENTS

This work was supported by the National Science Centre (NCN − Poland) Research Project: No. 2021/43/B/ST8/03207. The computational assistance was granted through the computing cluster GRAFEN at Biocentrum Ochota, the Interdisciplinary Centre for Mathematical and Computational Modelling of Warsaw University (ICM UW) and Poznań Supercomputing and Networking Center (PSNC).

## AUTHOR DECLARATION

Conflict of Interest

The author declare that they have no known competing financial interests or personal relationships that could have appeared to influence the work reported in this paper.

# Supplementary material
# Transferability of interatomic potentials for germanene (2D germanium)

Marcin Maździarz[a]

*Institute of Fundamental Technological Research Polish Academy of Sciences, Pawińskiego 5B, 02-106 Warsaw, Poland*

(Dated: 25 August 2023)

Crystallographic Information Files (CIF) for polymorphs of germanene (created by qAgate: Open-source software to post-process ABINIT):

- flat (F) germanene:

```
# Ge_F  P6/mmm (191)
data_Ge
_symmetry_space_group_name_H-M    'P  1'
_cell_length_a    4.06636944
_cell_length_b    4.06636944
_cell_length_c    20.00000009
_cell_angle_alpha    90.00000000
_cell_angle_beta    90.00000000
_cell_angle_gamma    120.00000000
_symmetry_Int_Tables_number    1
_chemical_formula_structural    Ge
_chemical_formula_sum    Ge2
_cell_volume    286.40084519
_cell_formula_units_Z    2
loop_
 _symmetry_equiv_pos_site_id
 _symmetry_equiv_pos_as_xyz
  1  'x, y, z'
loop_
 _atom_site_type_symbol
 _atom_site_label
 _atom_site_symmetry_multiplicity
 _atom_site_fract_x
 _atom_site_fract_y
 _atom_site_fract_z
 _atom_site_occupancy
   Ge   Ge0   1   0.33333333   0.66666667   0.50000000   1
   Ge   Ge1   1   0.66666667   0.33333333   0.50000000   1
```

- low-buckled (LB) germanene:

```
# Ge_LB  P-3m1 (#164)
data_Ge
_symmetry_space_group_name_H-M    'P  1'
_cell_length_a    3.95487807
_cell_length_b    3.95487807
_cell_length_c    20.00000009
```

---

[a] Electronic mail: mmazdz@ippt.pan.pl.

```
_cell_angle_alpha    90.00000000
_cell_angle_beta     90.00000000
_cell_angle_gamma    120.00000000
_symmetry_Int_Tables_number   1
_chemical_formula_structural   Ge
_chemical_formula_sum   Ge2
_cell_volume    270.91111689
_cell_formula_units_Z    2
loop_
 _symmetry_equiv_pos_site_id
 _symmetry_equiv_pos_as_xyz
  1  'x, y, z'
loop_
 _atom_site_type_symbol
 _atom_site_label
 _atom_site_symmetry_multiplicity
 _atom_site_fract_x
 _atom_site_fract_y
 _atom_site_fract_z
 _atom_site_occupancy
  Ge  Ge0  1  0.33333333   0.66666667   0.48409184   1
  Ge  Ge1  1  0.66666667   0.33333333   0.51590816   1
```

- trigonal dumbbell (TD) germanene:

```
# Ge_P3xP3td P-62m (#189)
data_Ge
_symmetry_space_group_name_H-M    'P 1'
_cell_length_a    6.69846671
_cell_length_b    6.69846671
_cell_length_c    20.00000009
_cell_angle_alpha    90.00000000
_cell_angle_beta     90.00000000
_cell_angle_gamma    120.00000000
_symmetry_Int_Tables_number   1
_chemical_formula_structural   Ge
_chemical_formula_sum   Ge7
_cell_volume    777.16178277
_cell_formula_units_Z    7
loop_
 _symmetry_equiv_pos_site_id
 _symmetry_equiv_pos_as_xyz
  1  'x, y, z'
loop_
 _atom_site_type_symbol
 _atom_site_label
 _atom_site_symmetry_multiplicity
 _atom_site_fract_x
 _atom_site_fract_y
 _atom_site_fract_z
 _atom_site_occupancy
  Ge  Ge0  1   0.00000000   0.00000000   0.50000000   1
  Ge  Ge1  1  -0.03334455   0.33333333   0.50000000   1
  Ge  Ge2  1   0.36667788   0.03334455   0.50000000   1
  Ge  Ge3  1   0.66666667   0.33333333   0.42557395   1
  Ge  Ge4  1   0.33333333   0.66666667   0.50000000   1
```

| | | | | | | |
|---|---|---|---|---|---|---|
| Ge | Ge5 | 1 | 0.66666667 | 0.63332212 | 0.50000000 | 1 |
| Ge | Ge6 | 1 | 0.66666667 | 0.33333333 | 0.57442605 | 1 |

- large honeycomb dumbbell (LHD) germanene:

```
# Ge_P3xP3lhd P6/mmm (191)
data_Ge
_symmetry_space_group_name_H-M    'P 1'
_cell_length_a    7.66720152
_cell_length_b    7.66720152
_cell_length_c    20.00000009
_cell_angle_alpha    90.00000000
_cell_angle_beta    90.00000000
_cell_angle_gamma    120.00000000
_symmetry_Int_Tables_number    1
_chemical_formula_structural    Ge
_chemical_formula_sum    Ge10
_cell_volume    1018.20303116
_cell_formula_units_Z    10
loop_
 _symmetry_equiv_pos_site_id
 _symmetry_equiv_pos_as_xyz
  1  'x, y, z'
loop_
 _atom_site_type_symbol
 _atom_site_label
 _atom_site_symmetry_multiplicity
 _atom_site_fract_x
 _atom_site_fract_y
 _atom_site_fract_z
 _atom_site_occupancy
  Ge  Ge0  1  0.18187276  0.36374553  0.50000000  1
  Ge  Ge1  1  0.36374553  0.18187276  0.50000000  1
  Ge  Ge2  1  0.18187276  0.81812724  0.50000000  1
  Ge  Ge3  1  0.33333333  0.66666667  0.42811425  1
  Ge  Ge4  1  0.66666667  0.33333333  0.42811425  1
  Ge  Ge5  1  0.81812724  0.18187276  0.50000000  1
  Ge  Ge6  1  0.63625447  0.81812724  0.50000000  1
  Ge  Ge7  1  0.81812724  0.63625447  0.50000000  1
  Ge  Ge8  1  0.33333333  0.66666667  0.57188575  1
  Ge  Ge9  1  0.66666667  0.33333333  0.57188575  1
```